\documentclass[structabstract]{aa}
\usepackage{natbib}
\bibpunct{(}{)}{;}{a}{}{,}
\usepackage{txfonts}
\usepackage{graphicx}
\usepackage{longtable}
\usepackage{lscape}
\title{Limits on the Orbits and Masses of Moons around Currently-Known
  Transiting Exoplanets}
\author{Carsten Weidner\inst{\ref{inst1}} \and Keith
  Horne\inst{\ref{inst1}}}
\institute{$^1$Scottish Universities Physics Alliance (SUPA), School of
  Physics and Astronomy, University of St. Andrews, North Haugh,
  St. Andrews, Fife KY16 9SS, UK
  \email{cw60,kdh1@st-andrews.ac.uk}\label{inst1}}

\begin{document}
\date{Received 2010 / Accepted 2010}

\label{firstpage}

\abstract{}
{Current and upcoming space missions may be able to detect moons of
  transiting extra-solar planets. In this context it is important to
  understand if exomoons are expected to exist and what their possible
  properties are.} 
{Using estimates for the stability of exomoon orbits from
  numerical studies, a list of 87 known transiting exoplanets is
  tested for the potential to host large exomoons.} 
{For 92\% of the sample, moons larger than Luna can be excluded
  on prograde orbits, unless the parent exoplanet's internal
  structure is very different from the gas-giants 
  of the solar system. Only WASP-24b, OGLE2-TR-L9, CoRoT-3b and
  CoRoT-9b could have moons above 0.4 $m_\oplus$, which is within the
  likely detection capabilities of current observational
  facilities. Additionally, the range of possible orbital radii of
  exomoons of the known transiting exoplanets, with two exceptions, is
  below 8 Jupiter-radii and therefore rather small.}
{}


\titlerunning{Moons around Hot-Jupiters}
\authorrunning{Weidner \& Horne}

\maketitle

\section{Introduction}
\label{se:intro}
Currently more than 460 exoplanets\footnote{For an up-to-date
  list see http://exoplanet.eu/} have been detected through
various methods. With improving instrument precision smaller
and less massive objects are or will be soon accessible
observationally. With current instruments like Kepler it should even be
possible to detect moons of exoplanets
\citep{SS99,SSS07,SSS09,K09,KFC09}. But as most planets are found by
methods most sensitive to massive planets with a small semi-major axis
(``Hot-Jupiters'') the question arises not only if it is possible to
detect exomoons but also how likely it is for them to form and survive in
the first place. Several studies \citep{BO02,DWY06} explore the
stability of orbits around gas giants.

In this contribution we apply to a sample of observed exoplanets the
results of \citet{DWY06} on the stability of moons around gas
giants. Our sample (Tab.~\ref{tab:exoplanets}) includes all published
transiting exoplanets for which the mass and radius of the planet and
the host star, and the orbital parameters are all reasonably well
known.

\section{Stability Domains for Exomoons}
\label{sec:setup}
The region of orbital stability around a close-in gas-giant planet is
set by two radii. We assume that the smallest orbit is set by the
Roche-radius. Any moon larger than a few km within the Roche-limit of
its planet would be torn apart by the tidal forces between the planet and
the moon. The Roche-radius, $R_\mathrm{roche}$, depends mainly on the
density of the two interacting objects and can be written for
fluid-like objects as \citep{BT87}:

\begin{equation}
\label{eq:roche}
R_\mathrm{roche} = 2.44~r_\mathrm{p}
\left(\frac{\rho_\mathrm{p}}{\rho_\mathrm{m}}\right)^{1/3} =
2.44~\left(\frac{4~\pi}{3}\right)^{-1/3}
\left(\frac{m_\mathrm{p}}{\rho_\mathrm{m}}\right)^{1/3},
\end{equation}
where $r_\mathrm{p}$ is the radius of the planet, $\rho_\mathrm{p}$
and $\rho_\mathrm{m}$ are the mean densities of the planet and the
moon, and $m_\mathrm{p}$ the mass of the planet. Of course the Roche
criterion also limits the minimal semi-major axis, $a$, of the planet's
orbit around its star. The second part of eq.~\ref{eq:roche}
shows that $R_\mathrm{roche}$ is independent of $r_\mathrm{p}$.

The outer limit for stable orbits of a moon around an exoplanet is
the so-called Hill-radius, which defines the sphere in which the
gravitational pull of the planet on the moon is larger than that of
the star. The Hill-radius is given as \citep{B86}: 

\begin{equation}
\label{eq:hill}
R_\mathrm{Hill} = a
\left(\frac{m_\mathrm{p}}{3~M_\mathrm{\ast}}\right)^{1/3},
\end{equation}
where $M_\mathrm{\ast}$ is the mass of the star.

By using numerical integrations of the equations of motion,
recent studies \citep[e.g.][]{BO02} found that the Hill-radius
over estimates the maximum stable orbital radius by a factor
$f$. \citet{DWY06} studied this question in detail and derived two
equations for the maximal stable orbital radii, one for prograde
motion of the moon and the other one for retrograde motion. Both
depend on the eccentricities, $e_\mathrm{p}$, for the planet's orbit
and $e_\mathrm{m}$ for the moon's. For a prograde satellite
\citet{DWY06} give:

\begin{equation}
\label{eq:hill_pro}
R_\mathrm{max, p} = R_\mathrm{Hill} \times 0.4895 (1 - 1.0305e_\mathrm{p} -
0.2738e_\mathrm{m})
\end{equation}
and for retrograde ones:

\begin{equation}
\label{eq:hill_retro}
R_\mathrm{max, r} = R_\mathrm{Hill} \times 0.9309 (1 - 1.0764e_\mathrm{p} -
0.9812e_\mathrm{m}).
\end{equation}

\citet{BO02} also studied the possible lifetime of a moon due to
orbital decay as a result of tidal dissipation of angular momentum.
Based on this result \citet{DWY06} also derived an equation for the
maximum mass a moon can have for a given distance to the planet:

\begin{equation}
\label{eq:moonmax}
m_\mathrm{m, max} = \frac{2}{13} (f\,R_\mathrm{Hill})^{13/2}
\frac{Q_\mathrm{p}}{3\,k_\mathrm{2P}\,T\,r_\mathrm{p}^{5}}
\sqrt{\frac{m_\mathrm{p}}{G}}.
\end{equation}
$Q_\mathrm{p}$ is the dimensionless tidal dissipation factor,
$k_\mathrm{2P}$ the tidal Love number, $T$ the moon's lifetime and
$f\,R_\mathrm{Hill}$ either the pro- or the retrograde 
Hill radius. $Q_\mathrm{p}$ is very poorly constrained even for the
planets of the solar system and even more uncertain for
exoplanets. Following \citet{BO02} we chose $Q_\mathrm{p}$ = 10$^5$
and $k_\mathrm{2P}$ = 0.51. For the satellite lifetime, $T$, we adopt
the minimum age of the parental star if given. For stars lacking
age determinations, a minimum age of 1 Gyr is used. We consider
moon/planet mass ratios $q \le$ 0.1, though, the tidal effects on the
planetary rotation by moons with $q >$ 0.01 might
already modify the result \citep{BO02}. Such cases are marked with a
'$d$' in Table~\ref{tab:exoplanets}.

The largest uncertainty in eq.~\ref{eq:moonmax} lies in
$Q_\mathrm{p}$. While a $Q_\mathrm{p} \sim$ 10$^5$ is commonly used,
\citet{CMA09} suggested values as high as 10$^{13}$ for
exoplanets. A recent study \citep{LAK09} derived $Q_\mathrm{p}$ =
3.6 $\times10^4$ for Jupiter through astrometric observations of the
planet and its moon Io. In this context it is 
interesting to note \citep{BO02} that the actual detection of exomoons
will give some important constrains on $Q_\mathrm{p}$, as
eq.~\ref{eq:moonmax} can be written the following way:

\begin{equation}
\label{eq:qp}
Q_\mathrm{P,min} = \frac{39}{2}\,(a_\mathrm{m})^{-13/2}\,
k_\mathrm{2P}\,m_\mathrm{m}\,T_\mathrm{min}\,r_\mathrm{p}^{5}
\sqrt{\frac{G}{m_\mathrm{p}}}. 
\end{equation}
Here only a minimal value for $Q_\mathrm{p}$ can be achieved as an
observed exomoon need not necessarily be the most-massive moon 
possible for that planet.

\section{Results}
\label{sec:res}
A list of currently known transiting exoplanets is shown in
Tab.~\ref{tab:exoplanets}. In addition to the observed parameters of
these planets, the table gives the Roche-radii, maximal pro- and
retrograde Hill-radii and the maximal pro- and retrograde moon masses
(eqs.~\ref{eq:roche}, \ref{eq:hill_pro}, \ref{eq:hill_retro} and
\ref{eq:moonmax}) for these systems. We used $Q_\mathrm{p}$ of
10$^5$ and $\rho_\mathrm{m}$ of 3 g cm$^{-3}$ for the calculations.

\begin{figure}
\begin{center}
\includegraphics[width=8cm]{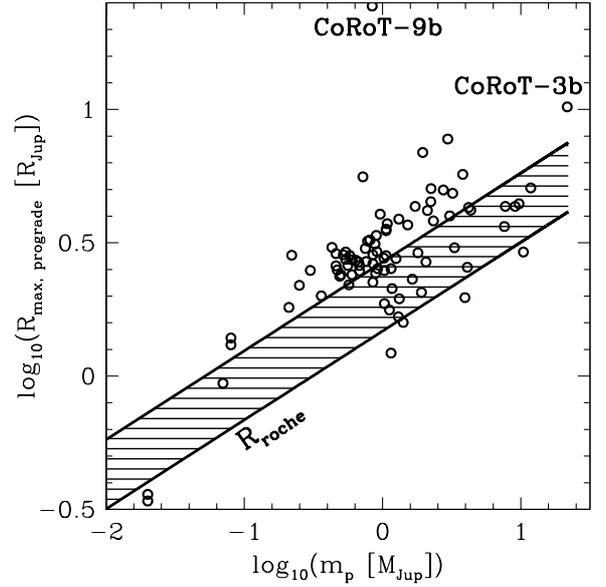}
\vspace*{-1.5cm}
\caption{Open circles: largest stable prograde orbital radii for
  moons of the known transiting exoplanets as shown in
  Tab.~\ref{tab:exoplanets}. The minimal stable orbits for any
  substantial moon is given by the Roche-radii for moons with
  $\rho_\mathrm{m}$ = 1 g cm$^{-3}$ and $\rho_\mathrm{m}$ = 6 g
  cm$^{-3}$ (shaded region). All moons are assumed to reside on
  circular orbits ($e_\mathrm{m}$ = 0).}
\label{fig:1}
\end{center}
\end{figure}

In Fig.~\ref{fig:1} the maximal stable prograde orbital radii,
$R_\mathrm{max, p}$,  for moons are shown for the known transiting
exoplanets from Tab.~\ref{tab:exoplanets}. Plotted as a shaded region
are the Roche limits for moons with densities between 1 and 6 g
cm$^{-3}$. For the majority of the known exoplanets stable moons on
prograde orbits are possible. Depending on the density of the moon the
percentage is between 63\% ($\rho_\mathrm{m}$ = 1 g cm$^{-3}$), 85\% 
($\rho_\mathrm{m}$ = 3 g cm$^{-3}$) and 93\% ($\rho_\mathrm{m}$ 
= 6 g cm$^{-3}$). 

It should be noted here, that the majority of the Hill-radii
derived through eq.~\ref{eq:hill_pro} agree within better than 10\%
with the ones derived by \citet{Do10} for the 43 exoplanets of their
sample which coincide with our sample. Though, \citet{Do10} does not
investigate the possible masses of the exomoons and does not consider
the Roche radii as an inner limits of the orbits of the moons.

\begin{figure}
\begin{center}
\includegraphics[width=8cm]{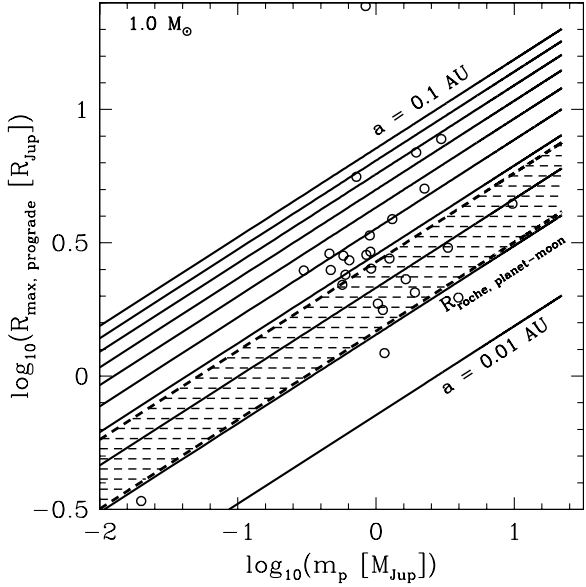}
\vspace*{-1.5cm}
\caption{Solid lines: Maximal stable prograde orbital radii vs
  planet mass for different semi-major axis of the planets for a 1.0
  $M_\odot$ star. The dashed-shaded region is minimal stable orbit
  given by the Roche-radii for moons with a density of
  $\rho_\mathrm{m}$~=~1 to 6 g cm$^{-3}$. For both the planet and the
  moon zero eccentricity is assumed. Open circles are known exoplanets
  orbiting a 1.0 $\pm$ 0.1 $M_\odot$ stars.}
\label{fig:2}
\end{center}
\end{figure}

The limiting $R_\mathrm{max, p}$, calculated from
eq.~\ref{eq:hill_pro} are shown as solid lines in Fig.~\ref{fig:2} for
exoplanets with planetary masses between 0.01 and 40
$M_\mathrm{Jupiter}$ around a 1.0 $M_\odot$ star for ten different
orbital separations, $a$, from 0.01 to 0.1 AU. The dashed-shaded
region marks the Roche-radii of the planet in respect to moons with
densities between 1 and 6 g cm$^{-3}$. The eccentricity for both the
planet and the moon is set to zero as non-zero eccentricities would
only reduce the Hill-radii. Moons around planets on orbits of
0.02 AU and less are excluded, while only high-density
($\rho_\mathrm{m} >$ 3 g cm$^{-3}$) moons can survive for planets on
the 0.03 AU orbit. Also plotted in the Fig.~\ref{fig:2} are the
exoplanets from Tab.~\ref{tab:exoplanets} for which the host stars are
within 0.1 $M_\odot$ of 1.0 $M_\odot$. 11 out of 27 of these
exoplanets have $R_\mathrm{max, p}$ within their Roche-radii,
depending on the density of the moon. Therefore, a detection of a moon
around one of these exoplanets would give strong constrains on the
density of the moon.

\begin{figure}
\begin{center}
\includegraphics[width=8cm]{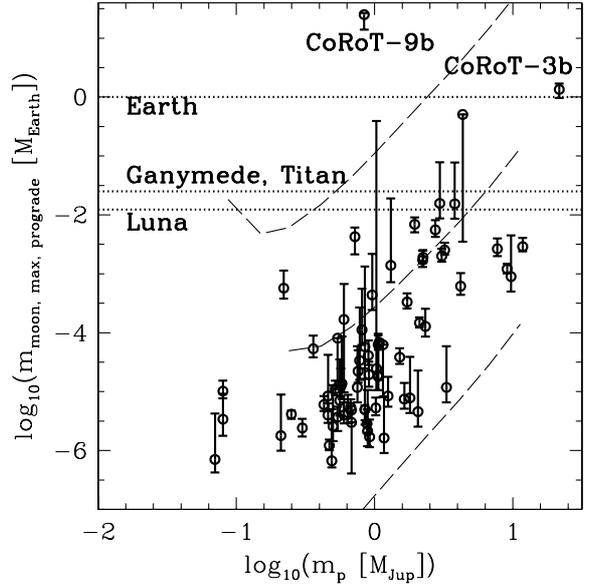}
\vspace*{-1.5cm}
\caption{Open circles: Maximal possible moon mass for prograde
  moons of the known transiting exoplanets from
  Tab.~\ref{tab:exoplanets}. Shown as horizontal dotted lines are the
  mass of the Earth, of Ganymede and Titan, the largest moons in the 
  solar system, and of the Earth's moon. For the error bars only the
  minimal and maximal ages of the stars are considered. All moons
  are assumed to reside on circular orbits ($e_\mathrm{m}$ =
  0). The dashed lines are the maximal possible moon masses for
    theoretical mass-radius relations for gaseous exoplanets from
    \citet{FMB07} around a 1 $M_\odot$ star. The top-most dashed line
    corresponds to 300 Myr old exoplanets on a 0.1 AU orbit and with a
    core mass, $m_\mathrm{core}$, of 25 $M_\oplus$. The parameters for
    the exoplanets on the dashed line in the middle are age = 1 Gyr,
    $a$ = 0.045 AU and $m_\mathrm{core}$ = 50 $M_\oplus$. The lower
    dashed line plots 4.5 Gyr old exoplanets with $a$ = 0.02 AU and
    without a solid core.}
\label{fig:3}
\end{center}
\end{figure}

The maximal possible masses for moons on prograde orbits around the
known transiting exoplanets (eq.~\ref{eq:moonmax}) are shown in
Fig.~\ref{fig:3}. As is visible only very few exoplanets (7 of 87,
8\%) have the potential for moons as massive as the Earth's moon or
larger, independent of the density of the moons. But as
eq.~\ref{eq:moonmax} scales linearly with $Q_\mathrm{p}$ of the
planet, larger moons would be possible for exoplanets with very
different internal structures than our solar system Gas Giants.
Also shown as dashed lines in Fig.~\ref{fig:3} are maximal moon
masses for theoretical mass-radius relations from \citet{FMB07} for
different ages of the exoplanets, different orbital separations and
different solid core fractions of the exoplanets. The top-most dashed
lines corresponds to 300 Myr old exoplanets with 25 $M_\oplus$ solid
cores, orbiting at 0.1 AU around a 1 $M_\odot$ star. The middle dashed
line shows explanets with 50 $M_\oplus$ core mass, which are 1 Gyr old
and at a distance of 0.045 AU to the host star. And finally the lowest
dashed line are 4.5 Gyr old exoplanets with $a$ = 0.02 AU and without
a solid core. In all three cases a host star mass of 1 $M_\odot$ is
assumed in order to calculate the Hill radii (eq.~\ref{eq:hill})
needed for eq.~\ref{eq:moonmax}.

Relatively small changes of the exoplanetary radii due to contraction
with time translate into a large changes of the maximal moon mass as 
eq.~\ref{eq:moonmax} depends to the one over 5th power on the
exoplanet radius.

\begin{figure}
\begin{center}
\includegraphics[width=8cm]{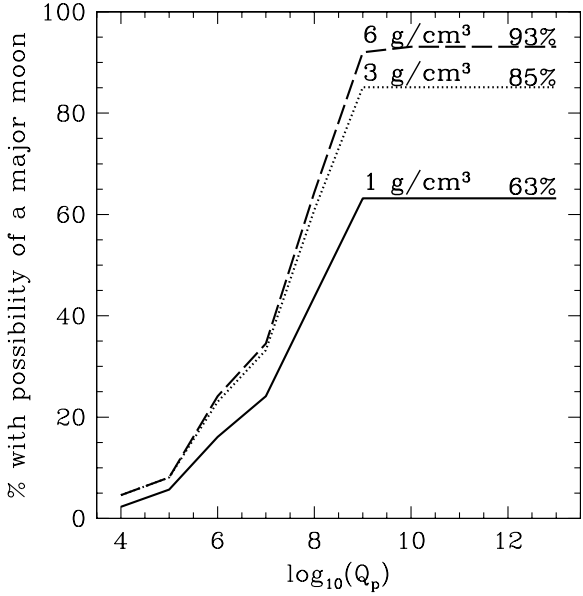}
\vspace*{-1.5cm}
\caption{Dependence of the fraction of exoplanets from
  Tab.~\ref{tab:exoplanets} with possible major moons
  ($m_\mathrm{exom} \ge m_\mathrm{Luna}$) on the tidal dissipation 
  factor, $Q_\mathrm{p}$, of the planet. The solid line refers to moon
with a density of 1 g cm$^{-3}$, the dotted line to 3 g cm$^{-3}$ and
the dashed line to 6 g cm$^{-3}$, as also indicated within the
plot. The final percentages reached are the fractions of planets which
can have any moon at all at a given density, which is independent of
$Q_\mathrm{p}$.}
\label{fig:results}
\end{center}
\end{figure}

The dependence of the fraction of exoplanets with possible major moons
on the tidal dissipation factor ($Q_\mathrm{p}$) is quantified in
Fig.~\ref{fig:results}. The fraction rises quite steeply
for $Q_\mathrm{p}$ values from 10$^4$ to 10$^9$ and then saturates at
the fraction of exoplanets which can have moons at all. This fraction
is set by the density of the moon through the Roche limit of the
orbit.

\section{Discussion \& Conclusions}
\label{se:dis}
With the use of the results of \citet{DWY06} on stable orbits around
gas giants the maximal and minimal orbital radii for hypothetical 
exomoons around the known transiting exoplanets are calculated.

Due to their much larger Hill-radii (for zero eccentricity of the
planet and the moon) retrograde moons have larger maximal possible
orbital radii and larger maximal masses than prograde moons.

Only WASP-24b \citep{SSB10}, OGLE2-TR-L9 \citep{SKV09,LAK10}, CoRoT-3b
\citep{DDA08} and CoRoT-9b \citep{DME10} can have a large
earth-sized (0.39, 0.51, 1.7 and 27 $M_\oplus$, respectively)
prograde moons, for a $Q_\mathrm{p}$ of 10$^5$. In the case of
WASP-24b and OGLE2-TR-L9 this is due to young minimum age of the
stars, for CoRoT-3b the large mass of the planet (brown dwarf) of about 22
$M_\mathrm{Jup}$ enables massive moons, and in the case of CoRoT-9b
the rather large distance of the planet to its parental star allows
for large Hill-radii and therefore large moons. Therefore, these four
exoplanets might be considered the best current candidates to search for
exomoons.

Additionally, SWEEPS-04 \citep{SCB06}, CoRoT-6b \citep{FHA10} and
CoRoT-13b \citep{CBO10} have
the potential for moons of the size of the Earth's moon
($m_\mathrm{Luna} \approx$ 1.2$\times10^{-2}$ $M_\oplus$). When
considering also retrograde moons several exoplanets could have
Earth-sized or even larger moons. But the formation of large
retrograde moons seems to be unlikely at least from the point of view
of the Solar System moons. Therefore, 92\% of the here studied
exoplanets probably can have only very small prograde moons. 

For WASP-19b \citep{HCT10}, CoRoT-7b \citep{QBM09}, WASP-18b
\citep{HAC09,SHD09}, WASP-12b \citep{HCL09}, OGLE-TR-56b \citep{UZS02,S10},
WASP-33b \citep{CGS10}, TrES-3 \citep{OCB07,S10}, WASP-4b \citep{WGHM08,S10},
OGLE-TR-113b \citep{BPS04,S10}, CoRoT-1b \citep{BBA08,PEC10},
CoRoT-14b \citep{TEG10} and GJ 1214b \citep{CBI09} prograde moons larger  
than a few kilometers radius are excluded. This translates into 15\%
of the total sample. Even when considering moons of a high density of
6 g cm$^{-3}$ and an extreme $Q_\mathrm{p}$ of 10$^{13}$, WASP-19b,
CoRoT-7b, WASP-18b and WASP-12b are excluded to have large moons.
In the case of HD 80606b \citep{NLM01,HTM10,HDD10} both retro- and
prograde moons are excluded due to the large eccentricity of the
planets orbit. For this planet, only for a moon with a density larger
than 31 g cm$^{-3}$ would be the Roche-radius smaller than the
prograde Hill-radius.

Therefore it is reasonable to conclude that for the vast majority
of transiting exoplanets the existence of exomoons as large as the
Earth moon or larger is rather unlikely, unless the tidal dissipation
factor $Q_\mathrm{p}$ is considerably larger than 10$^5$ for these
planets. The actual detection of exomoons would give considerable
insight into the internal structure of the host planet and possibly
the moon itself due to the relatively limited parameters space of
possible orbits around the known transiting exoplanets.

As the ``habitable zone'' of M dwarfs is very close to the star,
the rotational period of a planet becomes tidally locked to its
orbital period \citep{P77} or the planet could be in a spin-orbit
resonance, like Mercury in the Solar System. Several studies
\citep{JHR97,HDJ99,MS10} argue that despite tidal-locking such planets
might still bear life, though the odds for Earth-like life would seem
rather low. Exomoons around tidally-locked gas giants may circumvent
the problem of tidal-locking \citep{Ka10}. If close-in exoplanets or
the exomoons of Hot-Jupiters could be habitable (for Earth-like life)
is controversially discussed in the literature
\citep{P77b,WKW97,HDJ99,TBM07,KRL07,L07,KST07,SKS07,Ka10,CGK10,SWM10,JS10},
but of great interest as M dwarfs are the most common stars in the Galaxy.

The lowest-mass star in the current sample of transiting
exoplanets is the M4.5 dwarf GJ 1214, with a mass of $M_\mathrm{GJ
1214} \approx$ 0.16 $M_\odot$. A hypothetical Jupiter-sized gas
giant in the habitable zone of this star ($R_\mathrm{habit} \approx $
0.057 AU)\footnote{since from $T_\ast = L_\ast^{1/4}R_\ast^{-1/2}$ (in
  solar units) and $T_\mathrm{p} =  T_\ast
  \left(\frac{1-A_\mathrm{p}}{4}\right)^{1/4} 
  \left(\frac{R_\mathrm{\ast}}{a_\mathrm{p}}\right)^{1/2}$, follows
  $R_\mathrm{habit} \approx 1
  AU\,\left(\frac{L_\ast}{L_\odot}\right)^{1/2}\,
  \left(\frac{1-A_\mathrm{p}}{0.7}\right)^{-1/2}$.} 
could host a prograde moon no larger than the Earth moon, unless the
$Q_\mathrm{p}$ of the planet is significantly larger than 10$^5$ 
and any such moon would be very close to the planet ($R_\mathrm{max,pro}$
$\approx$ 8 $R_\mathrm{Jupiter}$).

It should also be noted here that 95\% of the exoplanets included here
have maximal possible prograde orbital radii less than 4$\times10^5$ km
($\approx$ 5.7 $R_\mathrm{Jupiter}$), independent of the density of
the moon and the $Q_\mathrm{p}$ of the planet. In our Solar System
only two moons with masses similar to that of Earth's moon are so
close to their planets: Jupiter's moon Io 
($m_\mathrm{Io}$ $\approx$ 1.2 $m_\mathrm{Luna}$, $a_\mathrm{Io}$
$\approx$ 6 $R_\mathrm{Jupiter}$) and the Earth moon itself
($a_\mathrm{Luna}$ $\approx$ 5.5 $R_\mathrm{Jupiter}$). The formation
of very massive moons within the Hill-sphere of close-in exoplanets
might therefore be considered very difficult \citep{Nf10} - at least
on the basis of our current knowledge of the Solar System. Even
if the formation of close-in massive exomoons is possible, the
contraction of its host planet with time \citep{FMB07} will lead to
the orbital decay and eventual destruction of less and less massive
exomoons over time due to the strong dependence of the maximal
possible moon mass (eq.~\ref{eq:moonmax}) on the radius of the planet.

\begin{acknowledgements}
We like to thank the referee Jason Barnes for helpful suggestions.
CW is happy to thank Christine Liebig, Moira Jardine, and Andrew
Collier Cameron for helpful discussions. The authors also gratefully
acknowledge the use of the Extrasolar Planets Encyclopedia
(http://exoplanet.eu/).
\end{acknowledgements}

\bibliographystyle{aa}
\bibliography{mybiblio}

\clearpage \onecolumn
\begin{landscape}
\begin{longtable}{ccccrccrcrrrcc}
\caption{\label{tab:exoplanets} A list of 87 transiting exoplanets from
  exoplanet.eu. For WASP-1 to 19 the stellar data is updated according
  to \citet{ECP10}. The mass ($M_\mathrm{\ast}$) and radius
  ($R_\mathrm{\ast}$) of the host star are in solar units, the age in
  Gyr, the orbital period of the planet in days, the semi-major axis
  ($a$) in astronomical units, the eccentricity of the orbit, the mass
  ($M_\mathrm{pl}$) and the radius ($R_\mathrm{pl}$) of the planet in
  Jupiter units are directly from the web-page. The other columns
  are derived here. These are the maximal orbital radius for
  retrograde moons ($R_\mathrm{max, r}$) and prograde ones
  ($R_\mathrm{max, p}$), the Roche-radius of the planet
  ($R_\mathrm{roche}$, for $\rho_\mathrm{m}$ = 3 g cm$^{-3}$) and
  the maximal mass of the moon for retrograde
  ($m_\mathrm{sat,max, r}$) and prograde ($m_\mathrm{sat,max, p}$)
  orbits for $Q_\mathrm{p}$ = 10$^5$ and the minimum age of the
  star. The additional radii are in Jupiter radii while the maximal
  masses are in Earth masses.}\\
\hline\hline
Name &$M_\mathrm{\ast}$ &$R_\mathrm{\ast}$&Age&period &$a$
&$e_\mathrm{pl}$\tablefootmark{a} &$M_\mathrm{pl}$ &$R_\mathrm{pl}$
&$R_\mathrm{max, r}$ & $R_\mathrm{max, p}$
&$R_\mathrm{roche}$&$m_\mathrm{m, max, r}$&$m_\mathrm{m, max, p}$\\
&$M_\odot$ &$R_\odot$ &Gyr&days &AU & &$M_\mathrm{Jup}$ &$R_\mathrm{Jup}$
  &$R_\mathrm{Jup}$ & $R_\mathrm{Jup}$ &$R_\mathrm{Jup}$
&$M_\oplus$&$M_\oplus$\\
\hline
\endfirsthead
\caption{continued.}\\
\hline\hline
Name &$M_\mathrm{\ast}$ &$R_\mathrm{\ast}$&Age&period &$a$
&$e_\mathrm{pl}$\tablefootmark{a} &$M_\mathrm{pl}$ &$R_\mathrm{pl}$
&$R_\mathrm{max, r}$ & $R_\mathrm{max, p}$
&$R_\mathrm{roche}$&$m_\mathrm{m, max, r}$&$m_\mathrm{m, max, p}$\\
&$M_\odot$ &$R_\odot$ &Gyr&days &AU & &$M_\mathrm{Jup}$ &$R_\mathrm{Jup}$
  &$R_\mathrm{Jup}$ & $R_\mathrm{Jup}$ &$R_\mathrm{Jup}$
&$M_\oplus$&$M_\oplus$\\
\hline
\endhead
\hline
\endfoot
WASP-19b&0.95&0.94& 0.6 $\pm$ 0.5 &  0.79&0.0164&0.02& 1.15&1.31& 2.33& 1.22& 1.95 & 1.1$\cdot10^{-4}$ & 0.0 \\
CoRoT-7b&0.91&0.82& 1.5 -0.3/+0.8 &  0.85&0.0172&0.00& 0.02&0.14& 0.66& 0.34& 0.50 & 2.3$\cdot10^{-5}$ & 0.0 \\
WASP-18b&1.28&1.23& 0.63 -0.53/+0.95 &  0.94&0.0205&0.01&10.43&1.17& 5.56& 2.92& 4.06 & 0.17 & 0.0 \\
WASP-12b&1.33&1.57& 5.0 $\pm$ 4.0\tablefootmark{b} &  1.09&0.0229&0.05& 1.41&1.79& 3.01& 1.59& 2.08 & 1.4$\cdot10^{-5}$ & 0.0 \\
OGLE-TR-56b&1.17&1.32& 2.0 -0.0/+4.0 &  1.21&0.0225&0.00& 1.30&1.20& 3.17& 1.67& 2.03 & 6.9$\cdot10^{-5}$ & 0.0 \\
WASP-33b&1.50&1.44& 0.025 -0.0/+0.475&  1.22&0.0256&0.00& $<$ 4.1\tablefootmark{c}&1.50& 4.87& 2.56& 2.98 & 5.2$\cdot10^{-2}$ & 0.0\\
TrES-3&0.92&0.81& 5.0 $\pm$ 4.0\tablefootmark{b} &  1.31&0.0226&0.00& 1.91&1.31& 3.92& 2.06& 2.31 & 4.3$\cdot10^{-4}$ & 0.0\\
WASP-4b&0.90&1.15& 5.0 $\pm$ 4.0\tablefootmark{b} &  1.34&0.0230&0.00& 1.12&1.42& 3.36& 1.77& 1.93 & 8.1$\cdot10^{-5}$ & 0.0\\
OGLE-TR-113b&0.78&0.77& 0.7 -0.0/+4.0 &  1.43&0.0229&0.00& 1.32&1.09& 3.71& 1.95& 2.04 & 8.9$\cdot10^{-4}$ & 0.0\\
CoRoT-1b&0.95&1.11& 5.0 $\pm$ 4.0\tablefootmark{b} &  1.51&0.0254&0.00& 1.03&1.49& 3.55& 1.87& 1.88 & 8.6$\cdot10^{-5}$ & 0.0\\
CoRoT-14b&1.13&1.21& 0.6 $\pm$ 0.2 &  1.51&0.0270&0.00& 7.60&1.09& 6.93& 3.64& 3.65 & 0.22 & 0.0 \\
GJ1214b&0.16&0.21& 6.0 -3.0/+4.0 &  1.58&0.0140&0.27& 0.02&0.24& 0.68& 0.36& 0.50 & 7.7$\cdot10^{-7}$ & 0.0\\
WASP-5b&0.99&1.01& 3.0 $\pm$ 1.4 &  1.63&0.0273&0.00& 1.64&1.17& 4.39& 2.31& 2.19 & 9.1$\cdot10^{-4}$ & 1.4$\cdot10^{-5}$\\
OGLE-TR-132b&1.26&1.34& 5.0 $\pm$ 4.0\tablefootmark{b} &  1.69&0.0306&0.00& 1.17&1.25& 4.06& 2.13& 1.96 & 5.3$\cdot10^{-4}$ & 8.2$\cdot10^{-6}$\\
CoRoT-2b&0.97&0.90& 5.0 $\pm$ 4.0\tablefootmark{b} &  1.74&0.0281&0.00& 3.31&1.47& 5.75& 3.03& 2.77 & 3.8$\cdot10^{-3}$ & 5.9$\cdot10^{-5}$\\
SWEEPS-11&1.10&1.45& 5.0 $\pm$ 4.0\tablefootmark{b} &  1.80&0.0300&0.00& 9.70&1.13& 8.43& 4.43& 3.96 & 0.29 & 4.5$\cdot10^{-3}$\\
WASP-3b&1.24&1.31& 5.0 $\pm$ 4.0\tablefootmark{b} &  1.85&0.0317&0.00& 2.06&1.45& 5.11& 2.68& 2.37 & 1.5$\cdot10^{-3}$ & 2.3$\cdot10^{-5}$\\
WASP-2b&0.84&0.83& 5.0 $\pm$ 4.0\tablefootmark{b} &  2.15&0.0314&0.00& 0.85&1.04& 4.29& 2.25& 1.76 & 1.6$\cdot10^{-3}$ & 2.5$\cdot10^{-5}$\\
HAT-P-7b&1.47&1.84& 5.0 $\pm$ 4.0\tablefootmark{b} &  2.20&0.0379&0.00& 1.80&1.42& 5.51& 2.90& 2.26 & 2.6$\cdot10^{-3}$ & 3.9$\cdot10^{-5}$\\
HD189733b&0.80&0.79& 0.6 -0.0/+4.0 &  2.22&0.0314&0.00& 1.15&1.15& 4.82& 2.53& 1.95 & 4.1$\cdot10^{-3}$ & 6.2$\cdot10^{-5}$\\
WASP-14b&1.32&1.30& 0.75 $\pm$ 0.25  &  2.24&0.0370&0.09& 7.72&1.26& 8.20& 4.33& 3.67 & 0.25 & 4.0$\cdot10^{-3}$\\
WASP-24b&1.13&1.15& 1.6 -1.6/+2.1 &  2.34&0.0359&0.00& 1.03&1.10& 4.73& 2.49& 1.88 & 26\tablefootmark{d} & 0.39\\
TrES-2&0.98&1.00& 5.1 $\pm$ 2.7  &  2.47&0.0356&0.00& 1.25&1.26& 5.25& 2.76& 2.00 & 1.2$\cdot10^{-3}$ & 1.8$\cdot10^{-5}$\\
OGLE2-TR-L9&1.52&1.53& 7.0$\cdot10^{-3}$ -0.0/+0.1 &  2.49&0.0413&0.00& 4.34&1.61& 7.97& 4.19& 3.03 & 33\tablefootmark{d} & 0.51\\
WASP-1b&1.27&1.49& 5.0 $\pm$ 4.0\tablefootmark{b} &  2.52&0.0382&0.00& 0.89&1.36& 4.61& 2.43& 1.79 & 7.0$\cdot10^{-4}$ & 1.1$\cdot10^{-5}$\\
XO-2b&0.98&0.96& 2.0 $\pm$ 1.0  &  2.62&0.0369&0.00& 0.57&0.97& 4.19& 2.20& 1.54 & 1.6$\cdot10^{-3}$ & 2.5$\cdot10^{-5}$\\
GJ436b&0.45&0.46& 6.0 -5.0/+4.0 &  2.64&0.0289&0.15& 0.07&0.37& 1.77& 0.94& 0.77 & 2.6$\cdot10^{-4}$ & 4.2$\cdot10^{-6}$\\
WASP-26b&1.12&1.34& 6.0 $\pm$ 2.0  &  2.76&0.0400&0.00& 1.02&1.32& 5.27& 2.77& 1.87 & 5.2$\cdot10^{-4}$ & 7.9$\cdot10^{-6}$\\
HAT-P-16b&1.22&1.24&2.0 $\pm$ 0.8& 2.77&0.0413&0.04&4.19&1.29& 8.15& 4.29& 3.00& 6.6$\cdot10^{-2}$& 1.0$\cdot10^{-3}$\\
HAT-P-5b&1.16&1.17& 2.6 $\pm$ 1.8  &  2.79&0.0408&0.00& 1.06&1.26& 5.38& 2.83& 1.90 & 3.8$\cdot10^{-3}$ & 5.8$\cdot10^{-5}$\\
CoRoT-12b&1.08&1.12& 6.30 $\pm$ 3.10 &  2.83&0.0402&0.07& 0.92&1.44& 4.79& 2.53& 1.81 & 2.1$\cdot10^{-4}$ & 3.3$\cdot10^{-6}$ \\
HD149026b&1.30&1.50& 2.0 $\pm$ 0.8  &  2.88&0.0429&0.00& 0.36&0.61& 3.80& 2.00& 1.32 & 5.8$\cdot10^{-3}$ & 8.9$\cdot10^{-5}$\\
HAT-P-3b&0.94&0.82& 0.4 -0.3/+6.5 &  2.90&0.0389&0.00& 0.60&0.89& 4.55& 2.40& 1.57 & 4.4$\cdot10^{-2}$ & 6.7$\cdot10^{-4}$\\
HAT-P-13b&1.22&1.56& 5.0 -0.8/+2.5 &  2.92&0.0426&0.02& 0.85&1.28& 5.02& 2.64& 1.76 & 3.8$\cdot10^{-4}$ & 5.9$\cdot10^{-6}$\\
CoRoT-11b&1.27&1.37& 2.00 $\pm$ 1.00 &  2.99&0.0436&0.00& 2.33&1.43& 7.26& 3.82& 2.46 & 1.7$\cdot10^{-2}$ & 2.6$\cdot10^{-4}$ \\
TrES-1&0.87&0.82& 2.5 $\pm$ 1.4  &  3.03&0.0393&0.00& 0.76&1.10& 5.11& 2.69& 1.70 & 3.3$\cdot10^{-3}$ & 5.0$\cdot10^{-5}$\\
HAT-P-4b&1.26&1.59& 4.2  0.6  2.6 &  3.06&0.0446&0.00& 0.68&1.27& 4.94& 2.60& 1.63 & 3.7$\cdot10^{-4}$ & 5.7$\cdot10^{-6}$\\
HAT-P-8b&1.28&1.58& 3.4 $\pm$ 1.0  &  3.08&0.0487&0.00& 1.52&1.50& 7.01& 3.69& 2.14 & 3.5$\cdot10^{-3}$ & 5.4$\cdot10^{-5}$\\
WASP-10b&0.75&0.70& 0.8 -0.2/+0.2 &  3.09&0.0371&0.06& 3.06&1.08& 7.57& 3.99& 2.70 & 0.17 & 2.7$\cdot10^{-3}$\\
OGLE-TR-10b&1.18&1.16& 1.1 -0.0/+7.0 &  3.10&0.0416&0.00& 0.68&1.72& 4.71& 2.48& 1.63 & 2.0$\cdot10^{-4}$ & 3.0$\cdot10^{-6}$\\
WASP-16b&1.00&0.94& 2.3 -2.2/+5.8 &  3.12&0.0421&0.00& 0.85&1.01& 5.42& 2.85& 1.76 & 8.6$\cdot10^{-2}$ & 1.3$\cdot10^{-3}$\\
XO-3b&1.21&1.38& 2.82 -0.82/+0.58 &  3.19&0.0454&0.26&11.79&1.22& 9.50& 5.08& 4.23 & 0.24 & 4.1$\cdot10^{-3}$\\
HAT-P-12b&0.73&0.70& 2.5 $\pm$ 2.0  &  3.21&0.0384&0.00& 0.21&0.96& 3.45& 1.81& 1.10 & 5.8$\cdot10^{-4}$ & 8.9$\cdot10^{-6}$\\
Kepler-4b&1.22&1.49& 4.5 $\pm$ 1.5  &  3.21&0.0456&0.00& 0.08&0.36& 2.50& 1.31& 0.80 & 1.0$\cdot10^{-3}$ & 1.5$\cdot10^{-5}$\\
Kepler-6b&1.21&1.39& 3.8 $\pm$ 1.0  &  3.23&0.0457&0.00& 0.67&1.32& 5.10& 2.68& 1.63 & 4.8$\cdot10^{-4}$ & 7.4$\cdot10^{-6}$\\
WASP-6b&0.83&0.85& 5.0 $\pm$ 4.0\tablefootmark{b} &  3.36&0.0421&0.05& 0.50&1.22& 4.55& 2.40& 1.48 & 8.3$\cdot10^{-4}$ & 1.3$\cdot10^{-5}$\\
WASP-28b&1.08&1.05& 5.0 -2.0/+3.0 &  3.41&0.0455&0.05& 0.91&1.12& 5.55& 2.93& 1.80 & 2.1$\cdot10^{-3}$ & 3.2$\cdot10^{-5}$\\
Kepler-8b&1.21&1.49& 3.84 $\pm$ 1.5  &  3.52&0.0483&0.00& 0.60&1.42& 5.20& 2.73& 1.57 & 4.3$\cdot10^{-4}$ & 6.6$\cdot10^{-6}$\\
HD209458b&1.00&1.15& 4.0 $\pm$ 2.0  &  3.52&0.0475&0.07& 0.64&1.38& 5.15& 2.72& 1.60 & 5.6$\cdot10^{-4}$ & 8.8$\cdot10^{-6}$\\
WASP-22b&1.11&1.13& 5.0 $\pm$ 4.0\tablefootmark{b} &  3.53&0.0468&0.02& 0.56&1.12& 4.94& 2.60& 1.53 & 2.3$\cdot10^{-3}$ & 3.5$\cdot10^{-5}$\\
Kepler-5b&1.37&1.79& 3.0 $\pm$ 0.6 &  3.55&0.0506&0.00& 2.11&1.43& 7.95& 4.18& 2.38 & 1.2$\cdot10^{-2}$ & 1.8$\cdot10^{-4}$\\
TrES-4&1.38&1.81& 4.7 $\pm$ 2.0  &  3.55&0.0509&0.00& 0.88&1.81& 5.96& 3.13& 1.78 & 3.2$\cdot10^{-4}$ & 5.0$\cdot10^{-6}$\\
OGLE-TR-211b&1.33&1.64& 5.0 $\pm$ 4.0\tablefootmark{b} &  3.68&0.0510&0.00& 0.75&1.26& 5.73& 3.01& 1.69 & 3.8$\cdot10^{-3}$ & 5.9$\cdot10^{-5}$\\
WASP-11/HAT-P-10b&0.82&0.75&11.2 $\pm$ 4.1  &  3.72&0.0439&0.00& 0.46&1.04& 4.92& 2.59& 1.43 & 4.1$\cdot10^{-4}$ & 6.3$\cdot10^{-6}$\\
WASP-17b&1.16&1.20& 3.0 -2.6/+0.9 &  3.74&0.0510&0.13& 0.49&1.74& 4.48& 2.37& 1.47 & 3.1$\cdot10^{-4}$ & 5.0$\cdot10^{-6}$\\
WASP-15b&1.19&1.48& 3.9 -1.3/+2.8 &  3.75&0.0499&0.00& 0.54&1.43& 5.21& 2.74& 1.51 & 3.6$\cdot10^{-4}$ & 5.5$\cdot10^{-6}$\\
WASP-25b&1.00&0.95& 2.5 $\pm$ 2.1  &  3.76&0.0474&0.00& 0.58&1.26& 5.38& 2.83& 1.55 & 5.6$\cdot10^{-3}$ & 8.6$\cdot10^{-5}$\\
HAT-P-6b&1.29&1.46& 2.3 -0.7/+0.5 &  3.85&0.0524&0.00& 1.06&1.33& 6.67& 3.51& 1.90 & 5.9$\cdot10^{-3}$ & 9.0$\cdot10^{-5}$\\
Lupus-TR-3b&0.87&0.82& 5.0 $\pm$ 4.0\tablefootmark{b} &  3.91&0.0464&0.00& 0.81&0.89& 6.16& 3.24& 1.73 & 3.6$\cdot10^{-2}$ & 5.6$\cdot10^{-4}$\\
HAT-P-9b&1.28&1.32& 1.6 -1.4/+1.8 &  3.92&0.0530&0.00& 0.78&1.40& 6.11& 3.21& 1.71 & 1.8$\cdot10^{-2}$ & 2.7$\cdot10^{-4}$\\
WASP-29b&0.82&0.85&13.0 -3.0/+5.0 &  3.92&0.0456&0.00& 0.25&0.74& 4.17& 2.19& 1.17 & 4.0$\cdot10^{-4}$ & 6.2$\cdot10^{-6}$\\
XO-1b&1.00&0.93& 4.5 -2.0/+2.0 &  3.94&0.0488&0.00& 0.90&1.18& 6.41& 3.37& 1.79 & 4.8$\cdot10^{-3}$ & 7.4$\cdot10^{-5}$\\
OGLE-TR-182b&1.14&1.14& 5.0 $\pm$ 4.0\tablefootmark{b} &  3.98&0.0510&0.00& 1.06&1.47& 6.77& 3.56& 1.90 & 6.2$\cdot10^{-3}$ & 9.6$\cdot10^{-5}$\\
OGLE-TR-111b&0.82&0.83& 1.1 -0.0/+7.0 &  4.01&0.0470&0.00& 0.53&1.07& 5.53& 2.91& 1.50 & 5.3$\cdot10^{-3}$ & 8.1$\cdot10^{-5}$\\
CoRoT-13b&1.09&1.01& 1.64 $\pm$ 1.52 &  4.04&0.0510&0.00& 1.31&0.89& 7.37& 3.88& 2.03 & 1.24 & 1.9$\cdot10^{-2}$ \\
CoRoT-5b&1.00&1.19& 6.9 $\pm$ 1.4  &  4.04&0.0495&0.09& 0.47&1.39& 4.73& 2.50& 1.45 & 9.7$\cdot10^{-5}$ & 1.5$\cdot10^{-6}$\\
XO-4b&1.32&1.55& 2.1 $\pm$ 0.6  &  4.13&0.0555&0.00& 1.72&1.34& 8.24& 4.33& 2.23 & 3.0$\cdot10^{-2}$ & 4.6$\cdot10^{-4}$\\
XO-5b&0.88&1.06& 8.5 $\pm$ 0.8  &  4.19&0.0487&0.00& 1.08&1.09& 7.09& 3.73& 1.91 & 4.9$\cdot10^{-3}$ & 7.6$\cdot10^{-5}$\\
SWEEPS-04&1.24&1.18& 5.0 $\pm$ 4.0\tablefootmark{b} &  4.20&0.0550&0.00& 3.80&0.81&10.86& 5.71& 2.90 & 5.0 & 7.7$\cdot10^{-2}$\\
CoRoT-3b&1.37&1.56& 2.0 -0.4/+0.8 &  4.26&0.0570&0.00&21.66&1.01&19.45&10.23& 5.18 & 110\tablefootmark{d} & 1.7\\
WASP-21b&1.01&1.06&10.0 -3.0/+4.0 &  4.32&0.0521&0.00& 0.30&1.07& 4.73& 2.49& 1.24 & 2.3$\cdot10^{-4}$ & 3.5$\cdot10^{-6}$\\
WASP-13b&1.03&1.34& 5.0 $\pm$ 4.0\tablefootmark{b} &  4.35&0.0527&0.00& 0.46&1.21& 5.48& 2.88& 1.43 & 2.8$\cdot10^{-3}$ & 4.2$\cdot10^{-5}$\\
HAT-P-1b&1.13&1.12& 3.6 $\pm$ 1.0  &  4.47&0.0554&0.07& 0.52&1.22& 5.40& 2.85& 1.49 & 9.8$\cdot10^{-4}$ & 1.5$\cdot10^{-5}$\\
HAT-P-14b&1.39&1.47& 1.3 $\pm$ 0.4  &  4.63&0.0606&0.11& 2.23&1.15& 8.54& 4.51& 2.43 & 0.16 & 2.5$\cdot10^{-3}$\\
Kepler-7b&1.35&1.84& 3.5 $\pm$ 1.0  &  4.89&0.0622&0.00& 0.43&1.48& 5.78& 3.04& 1.40 & 5.5$\cdot10^{-4}$ & 8.4$\cdot10^{-6}$\\
HAT-P-11b&0.81&0.75& 6.5 -4.1/+5.9 &  4.89&0.0530&0.20& 0.08&0.45& 2.62& 1.39& 0.80 & 5.6$\cdot10^{-4}$ & 9.2$\cdot10^{-6}$\\
WASP-7b&1.25&1.23& 5.0 $\pm$ 4.0\tablefootmark{b} &  4.95&0.0618&0.00& 0.96&0.92& 7.70& 4.05& 1.83 & 0.14 & 2.2$\cdot10^{-3}$\\
HAT-P-2b&1.36&1.64& 2.7 $\pm$ 0.5  &  5.63&0.0688&0.52& 9.09&1.16& 7.82& 4.33& 3.88 & 6.9$\cdot10^{-2}$ & 1.5$\cdot10^{-3}$\\
CoRoT-8b&0.88&0.77& 2.00 $\pm$ 1.00 &  6.21&0.0630&0.00& 0.22&0.57& 5.40& 2.84& 1.12 & 7.4$\cdot10^{-2}$ & 1.1$\cdot10^{-3}$ \\
WASP-8b&1.03&0.95& 4.0 $\pm$ 1.0  &  8.16&0.0801&0.31& 2.24&1.04& 9.40& 5.05& 2.43 & 0.14 & 2.5$\cdot10^{-3}$\\
CoRoT-6b&1.05&1.02& 5.0 $\pm$ 4.0\tablefootmark{b} &  8.89&0.0855&0.10& 2.96&1.17&14.66& 7.75& 2.67 & 5.0 & 7.9$\cdot10^{-2}$\\
CoRoT-4b&1.10&1.15& 1.0 -0.3/+1.0 &  9.20&0.0900&0.00& 0.72&1.19&10.63& 5.59& 1.67 & 0.40 & 6.1$\cdot10^{-3}$\\
HAT-P-15b&1.01&1.08&6.8 -1.6/+2.5& 10.86&0.0964&0.19&1.95&1.07& 12.99& 6.90& 2.32& 0.55& 9.1$\cdot10^{-3}$\\
CoRoT-10b &0.89&0.79& 1.98 $\pm$ 0.62 & 13.24&0.1055&0.53& 2.75&0.97& 8.98& 4.99& 2.60 & 0.37 & 8.1$\cdot10^{-3}$ \\
HD17156b&1.24&1.45& 3.06 -0.76/+0.64 & 21.22&0.1623&0.68& 3.21&1.02& 8.29& 4.85& 2.74 & 0.11 & 3.4$\cdot10^{-3}$\\
CoRoT-9b&0.99&0.94& 5.0 $\pm$ 4.0\tablefootmark{b} & 95.27&0.4070&0.11& 0.84&1.05&46.19&24.43& 1.75 & 27\tablefootmark{d} & 27\tablefootmark{d}\\
HD80606b&0.90&0.92& 7.63 $\pm$ 1.0  &111.44&0.4490&0.93& 3.94&0.92& 0.00& 1.97& 2.94 & 0.0 & 0.0\\
\end{longtable}
\tablefoot{
\tablefoottext{a}{The eccentricity is often fixed to zero and not a
  freely fitted parameter.}
\tablefoottext{b}{No age given in the literature. Therefore an age of
  5.0 $\pm$ 4.0 Gyr was assumed.}
\tablefoottext{c}{For this planet only an upper limit for the mass is
  available.}
\tablefoottext{d}{This value is probably not correct in detail as
  eq.~\ref{eq:moonmax} does not include tidal effects of the moon on
  the planet's rotation, which can be important for such a high
  mass-ratio of moon to planet.}
}
\end{landscape}

\label{lastpage}
\end{document}